# Micro moon versus macro moon: Brightness and size

D C Agrawal
Department of Farm Engineering
Banaras Hindu University
Varanasi 221005, India
Email: dca_bhu@yahoo.com

## Abstract

The moon, moonlight, phases of the moon and its relatively simple recurring cycle has been of interest since time immemorial to the human beings, navigators, astronomers and astrologers. The fact that its orbit is elliptical as well its plane is inclined with the plane of rotation of the earth gives rise to new moon to full moon and solar and lunar eclipses. During the phase of the full moon, the luminous flux and its apparent size will depend on its distance from the earth; in case it is at farthest point known as lunar apogee causes smallest full moon or micro full moon and if it is closest to us termed as lunar perigee will result in macro full moon – also known as super moon, a term coined by astrologer Richard Nolle in 1979.  The theoretical expressions for the lunar luminous fluxes on the earth representing the power of lunar light the earth intercepts in the direction normal to the incidence over an area of one square meter are derived for two extreme positions lunar apogee and lunar perigee. The expressions for the apparent sizes of full moons corresponding to said positions are also mentioned. It is found that full perigee moon is about 29% brighter and 14% bigger than the full apogee moon consistent with the reported values.
.

## Introduction

Plato had said, "Astronomy compels the soul to look upwards and leads us from this world to another". The Moon is a familiar sight to us all. From the time humans have walked the earth, they have observed with wonder the moon, moonlight, phases of moon, full moon and solar and lunar eclipses. But what really causes the appearance of the moon to change so drastically? The answer has to do with not just the position of the moon but the sun and the earth as well. The sun is centre of our solar system and the earth is revolving in an elliptical orbit with eccentricity of 0.0167 around the sun as the astronomical year goes by and rotating on its own axis. At the same time, moon is also revolving in an elliptical orbit with an average eccentricity of 0.0549 and time period of 29.53 days around the earth and rotating on its own axis (Fig.1). Changes in the position of the moon, earth and sun cause various phases of the moon, solar and lunar eclipses and tides in the oceans. The most early calendars were also based on lunar phases but even nowadays they are used for folkloristic (e.g. astrology) or religious (e.g. the Islamic or Hindu calendars) recurrences.

Half of the moon is always lit by the sun. As the moon orbits the earth, we see the different parts of the lighted area. The revolution of the moon around the earth makes the moon look as if it is changing shape in the sky. This is caused by different angles from which we see the lighted part of the moon's surface. These are called phases of the moon. The moon passes through many major shapes during a cycle that repeats itself every 29.53 days. The phases always follow one another in same order- new moon, waxing crescent, first quarter, waxing gibbous, full moon, waning gibbous, last quarter, waning crescent and then new moon once again (Fig. 2(a)). The illuminated part of the moon corresponding to these phases are respectively 0%, 25% on RHS, 50% on RHS, 75% on RHS, 100%, 75% on LHS, 50% on LHS, and 25% on LHS (Fig. 2(b)). The phase corresponding to full moon could be possible because the plane of moon revolution is inclined at an angle of 5.145 degree with plane of the rotation of the earth around the sun-the ecliptic plane; that is why lunar eclipses do not happen more often because during most months the moon is above or below the earth.

The moon provides a dim light to the earth during night by reflecting the Sun rays. The measure of this light is luminance which is defined in photometry as the total luminous flux or apparent intensity of light hitting or passing through a surface. It is analogous to the radiometric unit watts per square meter, but with the power at each wavelength weighted according to the luminosity function[1], a standardized model of human brightness perception. The SI unit of luminance is lux ($lx$) which is equivalent to one lumen per square meter. Everyone including the students and teachers of physics are fond of this moonlight and some attempts have been made in the past to determine its value theoretically and experimentally appearing in pedagogic journals[2,3], research journals[4], and Handbooks[5].

The luminous flux reaching the earth depends on the phase of the moon as well as the moon's albedo which is defined to be the ratio of the total amount of light the moon reflects to the amount of light incident upon it. At full moon, the earth, moon, and sun are in approximate alignment but the moon is on the opposite side of the earth, so the entire sunlit part of the moon is facing us. The moon and the earth are tidally locked in the sense that the same lunar surface always faces earth. The shadowed portion is entirely hidden from the view. The perceived size of the full moon and the luminous flux will also depend on its distance from the earth; in case it is at farthest point known as lunar apogee (Fig. 1) will give rise to smallest full moon- called micro full moon and if it is closest to us termed as lunar perigee will result in macro full moon – known as super moon. The word "super moon" was coined by the astrologer Richard Nolle[6] in 1979 describing a full moon at perigee-syzygy that is a moon which occurs with the moon at or near (within 90% of ) its closest approach to earth in a given orbit. The aim of the present paper is to estimate and compare the perceived sizes of these two moons and luminous fluxes from the micro full moon and super moon observed from the earth.

**Theory**
The solar energy is electromagnetic in nature which is characterized by wavelength $\lambda$, frequency $\nu$, and velocity $c$ satisfying the relation

$$c = \lambda\nu; 0 \leq \lambda \leq \infty, \infty \geq \nu \geq 0 \qquad (1)$$

The electromagnetic spectrum[7,8] extends from below the radio frequencies at the long-wavelength end through gamma radiation at the short-wavelength end covering wavelengths from thousands of kilometers down to a fraction of the size of an atom. Assuming that the Sun has as an uniform temperature $T$ over its surface the Planck's radiation law[8,9] says that

$$I(\lambda, T)d\lambda = \frac{\varepsilon(\lambda,T) \cdot A \cdot 2\pi hc^2 \cdot d\lambda}{\lambda^5[\exp(hc/\lambda kT)-1]} \text{ W} \qquad (2)$$

$I(\lambda, T)d\lambda$ is the power radiated between the wavelengths $\lambda$ and $\lambda + d\lambda$, $A$ is the surface area, $\varepsilon$ is the the emissivity and the constants $h$ and $k$, respectively, are Planck's constant and Boltzmann's constant. For simplicity, considering the Sun to be an ideal blackbody ($\varepsilon = 1$) the solar flux $Q$ emitted over all the wavelengths from the unit area ($A = 1\ m^2$) of the Sun is

$$Q = \int_0^\infty I(\lambda, T)d\lambda = \sigma T^4\ Wm^{-2} \qquad (3)$$

where $\sigma$ is the Stefan-Boltzmann constant. When this flux reaches the moon[10] or the earth this is diluted by the factor

$$f = R_S^2/d^2 \qquad (4)$$

giving rise to the value of solar constant as

$$S = \sigma T^4 \cdot f\ Wm^{-2} \qquad (5)$$

Here $R_S$ is the radius of the Sun and $d$ being the yearly mean distance between moon (or earth) and the Sun.

**Solar luminous constant on the moon (or earth).** It is well known that the wavelengths region $\lambda_i = 380$ nm to $\lambda_f = 760$ nm corresponds to the visible light; however the human eye is not equally sensitive to all wavelengths in this region. Rather its spectral efficiency is highest at wavelength $\lambda_m = 555$ nm and becomes vanishingly small outside this interval. This behavior is quantified by spectral luminous efficiency $V(\lambda)$ for photopic luminosity function which is plotted[11] in Fig. 3. Also, at wavelength $\lambda_m = 555$ nm the electromagnetic radiation of one watt provides a luminous flux of 683 lumens (L). The number 683 was once referred to as the "mechanical equivalent of light" in the literature[1]. Hence, according to (3) the luminous flux emitted at the surface of the Sun but with the power at each wavelength being weighted by multiplying it with $683V(\lambda)$ is given by

$$Q(\lambda_i \to \lambda_f) = \int_{\lambda_f}^{\lambda_i} \frac{683V(\lambda) \cdot 2\pi hc^2 \cdot d\lambda}{\lambda^5[exp(hc/\lambda kT)-1]} \qquad (6)$$

This is diluted by the factor[10] $f$ when it reaches the surface of the moon giving the value of Solar Luminous Constant On the Moon ($SLCOM$) as

$$SLCOM = Q(\lambda_i \to \lambda_f) \cdot f \qquad (7)$$

The above acronym variable $SLCOM$ represents the power of solar light arriving at right angle on the moon's surface covering an area of one square meter. Its unit is lux.

**Lunar luminous flux on the earth.** The moon will reflect the above flux (7) according to its albedo[12] $\kappa$ which when reaches the earth will be further diluted by the factor

$$g = R_m^2/\ell^2 \tag{8}$$

Here $R_m$ is the radius of the moon and $\ell$ is the distance between the moon and the earth. Multiplication of the expression (7) by moon's albedo $\kappa$ and the dilution factor $g$ gives the expression for the Lunar Luminous Flux On the Earth ($LLFOE$) as

$$LLFOE = SLCOM \cdot f \cdot \kappa \cdot g \tag{9}$$

Here the acronym variable $LLFOE$ represents the power of lunar light the earth intercepts in the direction normal to the incidence over an area of one square meter. Its unit is also lux. The condition that the earth intercepts in the direction normal to the incidence is normally fulfilled during full moon nights. The value of dilution factor $g$ according to expression (8) is inversely proportional to $\ell^2$ where $\ell$ being the distance between earth and the moon; this distance[13,14] varies between approximately 357,000 kilometers (222,000 mi) and 406,000 km (252,000 mi) due to its elliptical orbit around the earth. The mean distance ($\ell_{mean}$) between earth and the moon is 382,000 kilometers. In case a full moon occurs when the distance between earth moon system $\ell_{perigee}$ is close to around 357000 km will be termed as macro moon or super moon. However, when it occurs close to another extreme $\ell_{apogee}$ that is around 406000 km this is called micro moon. The dilution factor $g$ corresponding to these positions of the moon will be, respectively

$$g_{perigee} = R_m^2/\ell_{perigee}^2 \tag{10a}$$
$$g_{apogee} = R_m^2/\ell_{apogee}^2 \tag{10b}$$
$$g_{mean} = R_m^2/\ell_{mean}^2 \tag{10c}$$

The expressions for lunar luminous fluxes on the earth vide (9) will also have different values as follows

$$LLFOE_{perigee} = SLCOM \cdot f \cdot \kappa \cdot g_{perigee} \tag{11a}$$

$$LLFOE_{apogee} = SLCOM \cdot f \cdot \kappa \cdot g_{apogee} \qquad (11b)$$
$$LLFOE_{mean} = SLCOM \cdot f \cdot \kappa \cdot g_{mean} \qquad (11c)$$

These are the desired expressions of lunar luminous fluxes on the earth for the present paper which will be used below to derive the values of apparent brightness of the moons corresponding to the positions mentioned above.

**Apparent brightness of micro moon and macro moon.** A nearby flashlight may appear to be brighter than a distant streetlight, but in absolute terms if they are compared side by side the flashlight is much dimmer. This statement contains the essence of the problem of determining stellar brightness; optical astronomers almost always use something called the magnitude system[15,16] to talk about the brightness of stars or any other astronomical object. Magnitude ($M$) is the logarithmic measure of the brightness of an object, beyond the solar system, measured in a specific wavelength or pass band, usually in optical or near-infrared wavelengths. The brighter the object appears, the lower the value of its magnitude. The apparent magnitude, however is represented by a lower case letter $m$, of a celestial body in the solar system is a measure of its brightness as seen by an observer on earth, adjusted to the value it would have in the absence of the atmosphere. Generally the visible spectrum is used as a basis for the apparent magnitude. The apparent magnitude[15] of a body in the visible band having flux $F$ can be defined as,

$$m = -2.51 \, log_{10} \left( \frac{F}{2.56 \cdot 10^{-6}} \right) \qquad (12)$$

Here $2.56 \cdot 10^{-6}$ is the reference flux in the same band such that of Vega having apparent magnitude almost zero. The apparent magnitude of the moon corresponding to perigee and apogee locations will have following expressions

$$m_{perigee} = -2.51 \, log_{10} \left( \frac{LLFOE_{perigee}}{2.56 \cdot 10^{-6}} \right) \qquad (13a)$$
$$m_{apogee} = -2.51 \, log_{10} \left( \frac{LLFOE_{apogee}}{2.56 \cdot 10^{-6}} \right) \qquad (13b)$$

For the sake of completeness the value of apparent magnitude corresponding to mean position will also be computed from

$$m_{mean} = -2.51 \, log_{10}\left(\frac{LLFOE_{mean}}{2.56 \cdot 10^{-6}}\right) \qquad (13c)$$

These are the desired expressions for the apparent magnitudes which will be used in the numerical illustration.

**Apparent sizes of micro moon and super moon.** The apparent size of the full moon perceived by our eyes can be evaluated through the relation

$$h = Diameter\ of\ the\ moon(2R_M)/Its\ distance\ from\ the\ earth\ (\ell) \qquad (14)$$

The expressions for the apparent sizes in two extreme positions will be as follows.

$$h_{perigee} = 2R_M/\ell_{perigee} \qquad (15a)$$
$$h_{apogee} = 2R_M/\ell_{apogee} \qquad (15b)$$

Once again the perceived size of the moon corresponding to its mean position will also be calculated through the relation

$$h_{mean} = 2R_M/\ell_{mean} \qquad (15c)$$

This will be compared with the apparent size of the sun $H_{SUN}$ given by

$$H_{SUN} = Diameter\ of\ the\ sun(2R_S)/Its\ distance\ from\ the\ earth(d) \qquad (16)$$

to show that it coincides with the apparent size of the moon as noticed during the total solar eclipse. This will be discussed in the numerical work section.

**Parameterization of $V(\lambda)$.** This curve (Fig. 3) was parameterized by Agrawal, Leff and Menon[17] assuming a skewed Gaussian function

$$V_{approx}(\lambda) \approx exp(-az^2 + bz^3); z \equiv \lambda/\lambda_m - 1, \lambda_m = 555\ nm \qquad (17)$$
$$a = 87.868, b = 40.951, \chi^2 = 0.035 \qquad (18)$$

The above constants were obtained by using un-weighted least squares fit of 381 values[11] of $\ln[V(\lambda)]$. The values of $a$ and $b$ were re-examined[2] both by using the 39 values of $V(\lambda)$ in between 380-760 nm at an interval of 10 nm given in Table 1 of reference 17 as well as 381 values[11] at an interval of 1 nm.

There is practically no difference between these two cases and the better chi-square fit so obtained corresponds to the values

$$a = 88.90, b = 112.95, \chi^2 = 0.017 \qquad (19)$$

The curve corresponding to the above parameters overlaps with the experimental curve shown in figure 3 and therefore it has not been depicted.

**Numerical work**

The solution of the integral of Eq.(6) is not possible analytically therefore this was evaluated numerically by Simpson's rule in the wavelength region $\lambda_i = 380$ nm to $\lambda_f = 760$ nm and substituting[8,12,18] the values of

Radius of the Sun $(R_S) = 6.96 \cdot 10^8$ m,
Radius of the Moon $(R_M) = 1.74 \cdot 10^6$ m,
Distance between Moon/Earth and Sun $(d) = 1.5 \cdot 10^{11}$ m, $\qquad (20)$
Temperature of Sun's photosphere $(T) = 5776\ K$
Mean distance between Moon and Earth $(\ell_{mean}) = 3.82 \cdot 10^8$ m
Shortest distance between Moon and Earth $(\ell_{perigee}) = 3.57 \cdot 10^8$ m,
Longest distance between Moon and Earth $(\ell_{apogee}) = 4.06 \cdot 10^8$ m,
Albedo of the Moon $(\kappa) = 12\%$,

as well as the values of fundamental constants

Planck's constant $(h) = 6.63 \cdot 10^{-34}\ J \cdot s$,
Stefan Boltzmann Constant $(\sigma) = 5.67 \cdot 10^{-8} Wm^{-2}K^{-4}$, $\qquad (21)$
Boltzmann constant $(k) = 1.38 \cdot 10^{-23} JK^{-1}$,
Velocity of light $(c) = 3.0 \cdot 10^8\ ms^{-1}$

the dilution factors $f, g_{perigee}, g_{apogee}$ and $g_{mean}$ were estimated through expressions (4), (10a), (10b) and (10c), respectively.

$f = 2.153 \cdot 10^{-5}$ $\qquad$ (22a)
$g_{perigee} = 2.376 \cdot 10^{-5}$ $\qquad$ (22b)
$g_{apogee} = 1.837 \cdot 10^{-5}$ $\qquad$ (22c)
$g_{mean} = 2.075 \cdot 10^{-5}$ $\qquad$ (22d)

This eventually gives the final desired results of solar luminous constant on moon $SLCOM$ and lunar luminous fluxes on the earth $LLFOE_{perigee}$, $LLFOE_{apogee}$ and $LLFOE_{mean}$ corresponding to two extreme positions and one mean position of full moons through expressions (7), (11a), (11b) and (11c), respectively.

$$SLCOM = 122.7 \cdot 10^3 \, lx \qquad (23a)$$
$$LLFOE_{perigee} = 0.350 \, lx \qquad (23b)$$
$$LLFOE_{apogee} = 0.271 \, lx \qquad (23c)$$
$$LLFOE_{mean} = 0.305 \, lx \qquad (23c)$$

The luminous flux observed from the earth during super moon[19,20] is approximately 29% greater than that due to micro moon and around 15% more with respect to the mean position full moon.

It is worth converting the above fluxes into a better scale for brightness known as apparent magnitude[15,16] (12) which is a logarithmic scale as is our eye. The corresponding values $m_{perigee}, m_{apogee}$ and $m_{mean}$, respectively, are

$$m_{perigee} = -12.891 \qquad (24a)$$
$$m_{apogee} = -12.612 \qquad (24b)$$
$$m_{mean} = -12.741 \qquad (24c)$$

As far as apparent sizes of the two moons under discussion and the mean position full moon are concerned they can be evaluated via expressions (13a), (13b) and (13c) and the numbers are follows.

$$h_{perigee} = 0.00975 \qquad (25a)$$
$$h_{apogee} = 0.00857 \qquad (25b)$$
$$h_{mean} = 0.00911 \qquad (25c)$$

This shows that the apparent size of super moon[19,20] corresponding to perigee position is around 14% larger than micro moon and about 7% larger with respect to the mean position full moon. The perceived size of the sun can be computed from the expression (16). It comes out to be

$$H_{SUN} = 0.00928 \qquad (26)$$

This value is almost equal to the perceived size of full moon at mean position vide (25c) and this is the reason why the sun is fully covered during total solar eclipse (Fig. 4).

**Conclusions**

The moon is a natural satellite of the earth and displays a relatively simple recurring cycle from a new moon to a full moon. The orbit of moon being elliptical in nature the two extreme locations on the orbit are perigee and apogee; the first being closest to the earth and the second corresponds to farthest. The fact that moon's orbit plane is inclined by 5.145 degree to the plane of rotation of the earth we observe full moon when the moon, earth and the sun fall on almost in one straight line. The other possibilities are of either total lunar eclipse or total solar eclipse. In both the cases the sun, earth and moon are aligned exactly; while in case of total lunar eclipse the earth comes between the sun and the moon and casts a shadow on the full moon whereas during total solar eclipse the moon covers the sun completely. The aim of the present paper was to find and compare the expressions and values of luminous fluxes reflected by micro and macro full moons as well as their perceived sizes observed from the earth. The major conclusions of the present work are summarized below.

- The theoretical expressions for the solar luminous constant on the moon $SLCOM$ [cf. Eq.(7)] which represents the power of solar light arriving at right angle on the moon's surface covering an area of one square meter and lunar luminous flux on the earth $LLFOE$ [cf. Eq.(9)] representing the power of lunar light the earth intercepts in the direction normal to the incidence over an area of one square meter are derived for the first time for the benefit of students.
- The value of solar luminous constant on the moon comes out to be $SLCOM = 122.7 \cdot 10^3 \ lx$
- The reflected lunar luminous flux intercepted by the earth corresponding to two extreme positions, perigee and apogee, are $0.350 \ lx$ and $0.271 \ lx$, respectively. The first case corresponds to macro moon known as super moon while the second one is known as micro moon. These values suggest that super moon is 29% brighter than micro moon which is consistent with the quoted value of 30% in the literature[19,20].
- A better scale for brightness known as apparent magnitude (12) which is a logarithmic scale as is our eye has been discussed and the corresponding values are $m_{perigee} = -12.891, and \ m_{apogee} = -12.612$ , respectively.
- The apparent size of the super moon is shown to be 14% bigger than micro full moon and this also coincides with the reports in the literature[19,20].

- The perceived sizes of the sun and the moon are almost equal and that is why during total solar eclipse the sun is fully covered by the new moon.

It may be added that in ancient times all sorts of religious, mystical and magical manifestation were often linked with the phases of the moon which were eventually proved to be merely randomly associated. Belief in some lunar-related phenomena persist even today, as in astrology, menstrual cycle rhythm, mood swings, baby gender bias, crime, etc; people are still described as lunatics if displaying unusual behaviour. Astrology is still not a real science, but merely trying to make connections between astronomical and mystical events which are significantly correlated.

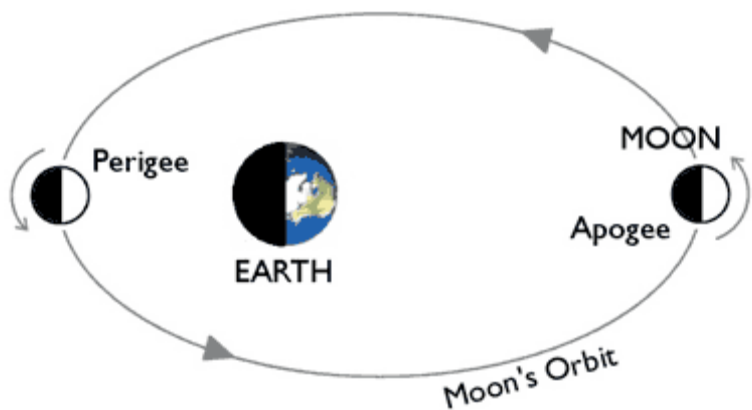

FIG. 1 The moon revolves in an elliptical orbit with an average eccentricity of 0.0549 and time period of 29.53 days around the earth and also rotates on its own axis; the figure corresponds to a rather higher eccentricity. The moon and the earth are tidally locked in the sense that the same lunar surface always faces the earth. The two extreme locations on the orbit are perigee and apogee; the first being closest to the earth and the second corresponds to farthest.

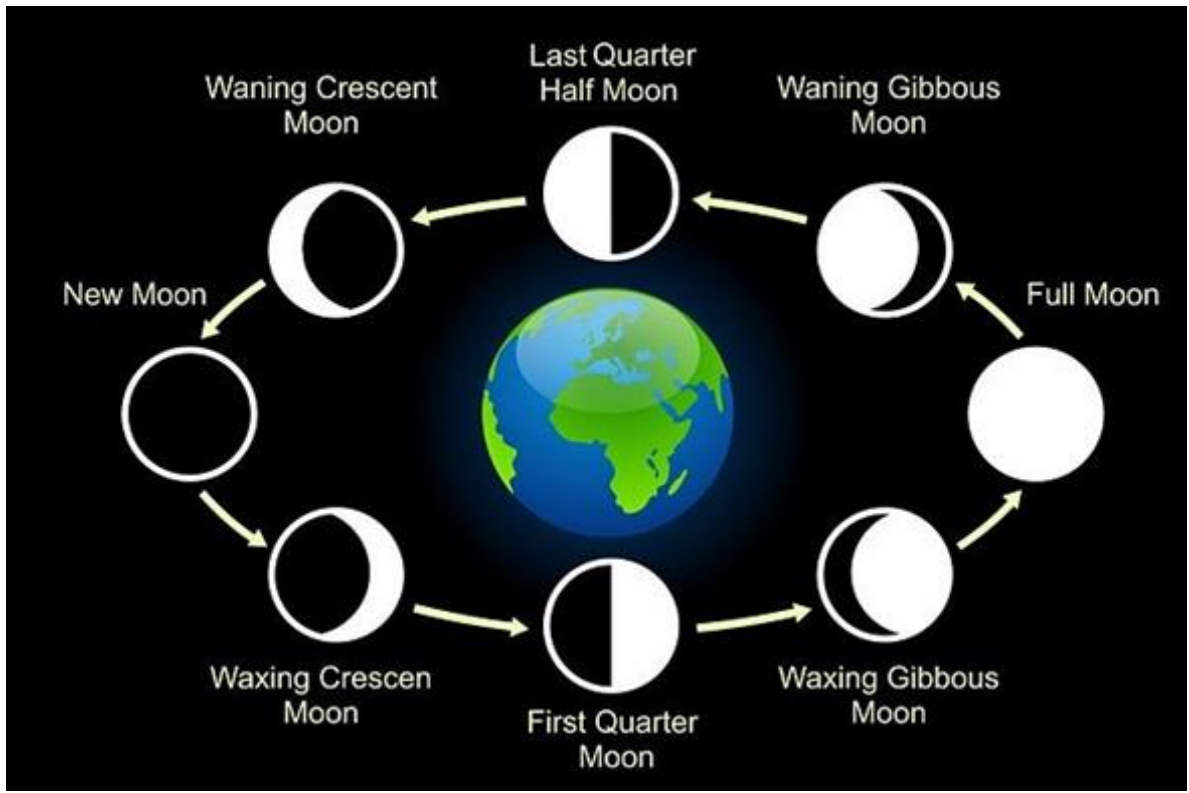

Fig. 2(a) Half of the moon is always lit by the sun. As the moon orbits the earth, we see the different parts of the lighted area. The revolution of the moon around the earth makes the moon look as if it is changing shape in the sky. This is caused by different angles from which we see the lighted part of the moon's surface. These are called phases of the moon. The moon passes through many major shapes during a cycle that repeats itself every 29.53 days. The phases always follow one another in same order- new moon, waxing crescent, first quarter, waxing gibbous, full moon, waning gibbous, last quarter, waning crescent and then new moon once again.

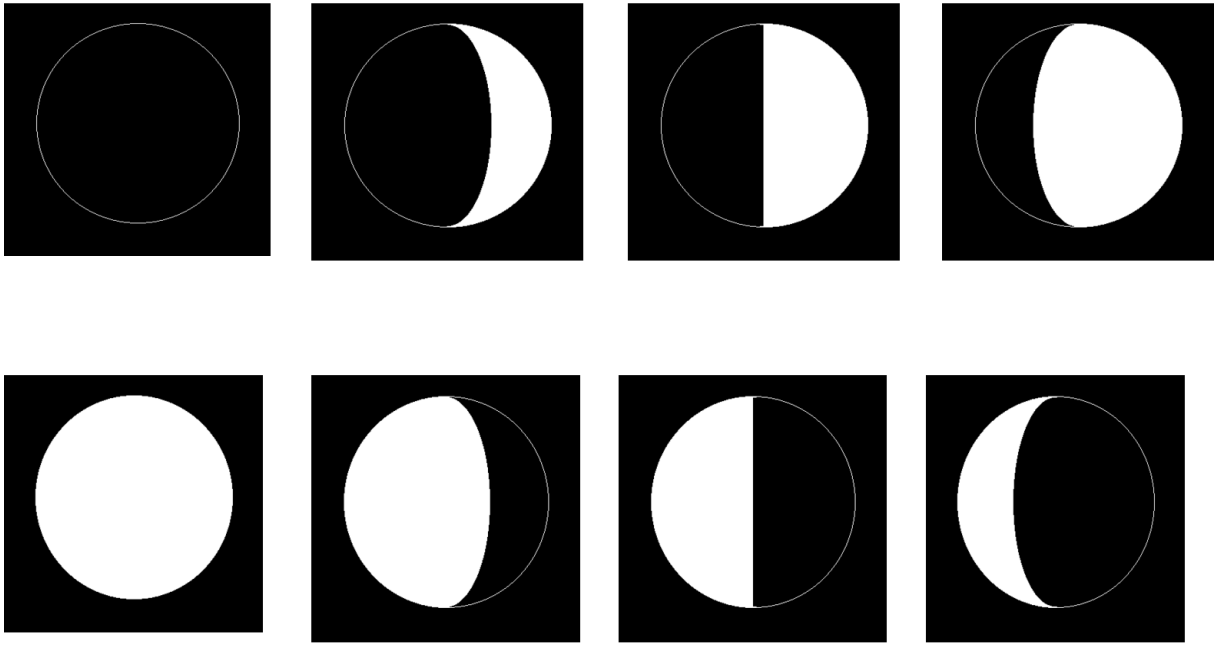

Fig 2(b). The illuminated part of the moon corresponding to the phases shown in Fig 2(a) are respectively 0%, 25% on RHS, 50% on RHS, 75% on RHS, 100%, 75% on LHS, 50% on LHS, and 25% on LHS.

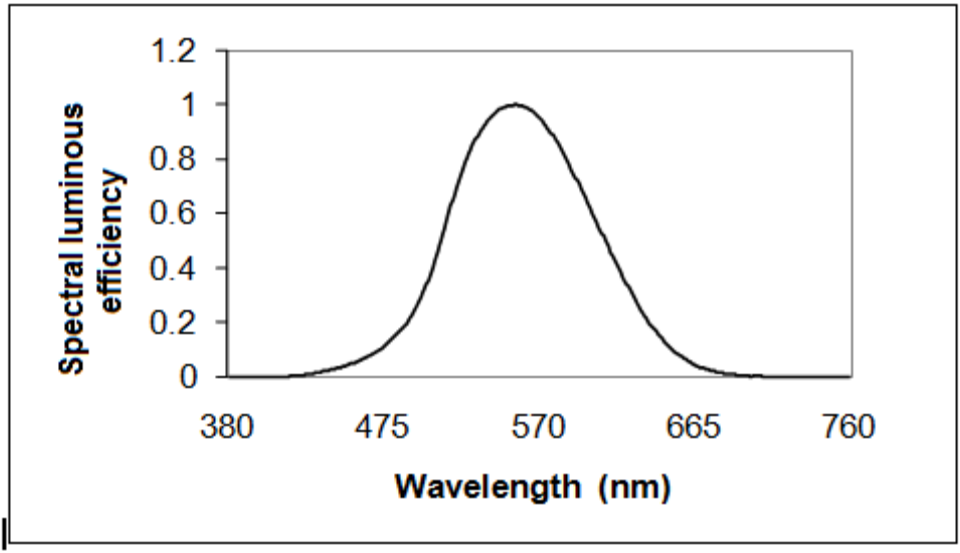

Fig. 3 Plot of the spectral luminous efficiency values[11] $V(\lambda)$ against the wavelength $\lambda$. It is well known that the wavelengths region $\lambda_i = 380$ nm to $\lambda_f = 760$ nm corresponds to the visible light; however the human eye is not equally sensitive to all wavelengths in this region. Rather its spectral efficiency is highest at wavelength $\lambda_m = 555$ nm and becomes vanishingly small outside this interval.

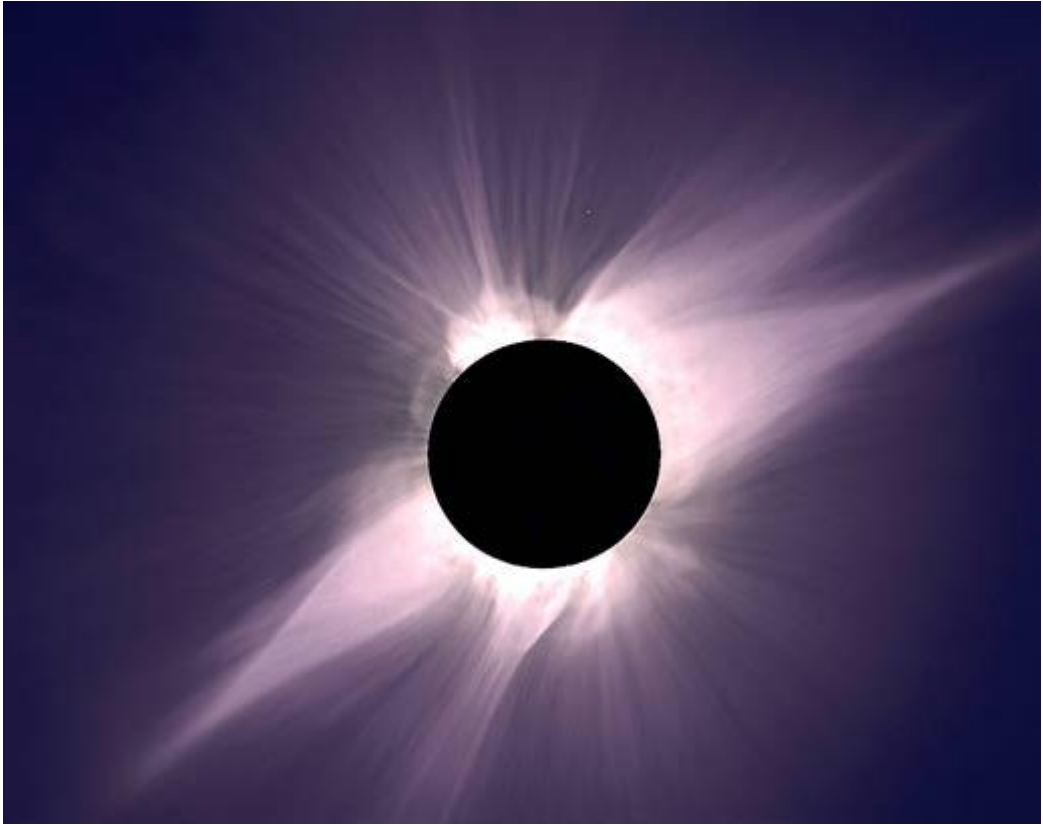

Fig. 4  A photograph of total solar eclipse. This happens while the moon comes in between the sun and the earth and covers the full sun. This is because the apparent sizes of sun and moon are almost equal [cf. (25c) and (26)].